%% file: main.tex
\documentclass[a4paper]{article}

\usepackage{INTERSPEECH2022}
\usepackage{cite}
\title{Using Active Speaker Faces for Diarization in TV shows}
\name{Rahul Sharma, Shrikanth Narayanan}
\address{
  University of Southern California, USA}
\email{rahul.sharma@usc.edu, shri@ee.usc.edu}
    
\begin{document}

\maketitle
\begin{abstract}
  Speaker diarization is one of the critical components of computational media intelligence as 
  it enables a character-level analysis of story portrayals and  media content understanding. Automated audio-based speaker diarization of entertainment media poses challenges due to the diverse acoustic conditions present in media content, be it background music, overlapping speakers, or sound effects. At the same time, speaking faces in the visual modality provide complementary information and not  prone to the errors seen in the audio modality. In this paper, we address the problem of speaker diarization in TV shows using the active speaker faces. We perform face clustering on the active speaker faces and show superior speaker diarization performance compared to the state-of-the-art audio-based diarization methods. We additionally report a systematic analysis of the impact of active speaker face detection quality on the diarization performance. We also observe that a moderately well-performing active speaker system could outperform the audio-based diarization systems. 
\end{abstract}

\input{intro}

\input{related_work}

\input{methods}

\input{Experiments}

\section{Conclusion}
This work compared the use of active speaker faces for speaker diarization in media content against conventional audio-only methods. We used an off-the-shelf active speaker face detection system 
and performed DBSCAN clustering, not requiring information about the number of characters. We demonstrated that this system outperforms the state-of-the-art speaker embeddings based diarization system. We further performed a detailed analysis of the impact of active speaker detection quality on speaker diarization performance. We observed that even moderately performing (~60\% accurate) active speaker systems can have sufficient information to outperform speaker embeddings based clustering for speaker diarization. 
Future work will develop methods to effectively combine face and speech information from talking characters for diarization. Another critical step is further improving the active speaker detection systems since face clustering showed high potential for speaker diarization. 
\newpage
\bibliographystyle{IEEEtran}

\bibliography{mybib}

\end{document}

%% file: intro.tex
\section{Introduction}
We use digital media in our daily lives to create, consume and experience stories in various forms: books, movies, TV shows, broadcast TV, etc. It impacts how we perceive, form, and communicate opinions and ideas. Recently there have been multiple efforts in the emerging field of computational media intelligence (CMI) to characterize and understand the impact of media portrayals on society. CMI focuses on developing computational tools to analyze  media content to obtain a holistic understanding of the stories being told and their impact on the experiences and behavior of individuals~\cite{somandepalli2021computational}. 

One of the central components of CMI is understanding the portrayals of characters in media along with their representation among the dimensions of character attributes such as age, gender, race, and appearance. It also aims to analyze the interactions between the characters. To address these aspects of CMI, knowing who speaks when in the media session is crucial. In this work, we address the broad problem of speaker diarization in dialogues of TV shows: determining the active speaker leveraging visual information. 

Speaker diarization refers to assigning all the speech segments in an audio session to their respective speakers, without using prior knowledge of the involved speakers, be their real identity or the number of participating speakers. There has been a plethora of work in diarization using the audio modality\cite{park2022review}. The conventional speaker diarization system involves voice activity detection and speaker turn detection to get speaker homogeneous speech segments, followed by clustering these segments in the speaker embedding space. Such solely audio-driven speaker diarization systems are prone to errors when applied to media content domain, consisting of narrative films, due to diverse and heterogeneous acoustic conditions and context: speech in background music, sound effects, overlapping speech, and wide variation in characters and their portrayals. 

Unlike read-speech as in broadcast news, in media such as TV shows and movies, character interaction is more spontaneous, thus leading to more frequent turn change and smaller speech segments, making audio-based diarization systems further prone to errors. However, at the same time, the visual modality contains substantial clues for speaker identification, especially the dialogue deliveries in entertainment media. 
In this work, we investigate the applicability of active speaker face clustering for supporting speaker diarization in media content; our exemplary use case is TV shows. 

We use a self-supervised framework followed by an audio-visual profile matching strategy, introduced in our previous works~\cite{sharma2020cross, sharma2022audio, sharma2019icip}, to obtain active speaker faces in the videos from TV episodes (Friends show). We take advantage of established face verification~\cite{2015Facenet}, and speaker recognition~\cite{2018speaker} features to represent the active speaker faces and the corresponding speech segment. To obtain diarization results, we cluster the face features using DBSCAN~\cite{DBSCAN} providing no information about the number of participating characters. In this work, we present experimental results on specific episodes of Friends, a part of Video Person Clustering Dataset (VPCD)~\cite{2021VPCD}. We are interested in exploring the use of active speaker faces for diarization and in comparing them against the speaker embeddings derived from speech segments. 

In this paper, our contributions are twofold: i) We show that active speaker face representations outperform state-of-the-art audio-based embeddings  for the task of speaker diarization in TV shows. Face representations are coupled with a simple DBSCAN-based clustering, thus avoiding sophisticated refinement procedures explicitly designed in recently proposed speaker diarization clustering schemes~\cite{wang2018speaker}. ii) We report a systematic analysis on the impact of the quality of active speaker detection systems on speaker diarization performance. We show that even a moderately well-performing active speaker detection system ($\sim$60\% accurate) can exceed the performance of audio-only systems.

%% file: related_work.tex
\section{Related Work}
\subsection{Active speaker detection}
Earlier works on active speaker localization in visual frames focused on detecting activity in the lip regions of the appearing faces in the frames~\cite{everingham2006hello}. More recently, ~\cite{roth2020ava} released a large-scale dataset consisting of active speaker annotations for parts movies, which was followed by the introduction of a wide variety of fully-supervised methods for active speaker localization~\cite{chung2019naver, alcazar2020active, zhang2019multi}. In parallel, there have been approaches posing the active speaker detection as a sound source localization task and proposed self-supervised systems trained for audio-visual synchrony in videos~\cite{owens2018audio, chung2017lip, 2021}. In our previous approaches, we proposed a weakly supervised cross-modal system trained for the presence of speech in videos and used the class activation maps for active speaker localization~\cite{sharma2019icip,sharma2020cross}. We further proposed audio-visual character profile matching to improve the performance of active speaker localization~\cite{sharma2022audio}. 

\subsection{Face and speaker recognition}
\textbf{Face recognition:} Approaches in face recognition can be divided into 2 categories: i) \emph{Face verification task}~\cite{lu2015surpassing, LFWTech} and ii) \emph{Face identification tas}k~\cite{parkhi2015deep,guo2016ms, guillaumin2009you}. A common approach to face recognition is metric learning~\cite{guillaumin2009you}, where a deep neural network~\cite{hu2018squeeze} is trained for the task of face verification or identification~\cite{guo2016ms, cao2018vggface2}.
\textbf{Speaker Recognition:} Speaker recognition is a well-explored field with earlier approaches focused on learning speaker embeddings using variants of the softmax classification loss~\cite{nagrani2020voxceleb, 2020InDefence, kenny2013plda, huang2018angular, yu2019ensemble}. Recent efforts developing metric learning objectives to learn an embedding space with small intra-class and large inter-class distances have shown promising results for speaker recognition~\cite{zhang2018text, wang2018speaker, 2020InDefence}.


\subsection{Speaker diarization}
Speaker diarization in audio modality has been extensively addressed, and an active area of research~\cite{park2022review}. In general, diarization frameworks consist of multistage paradigms involving voice activity detection, speaker embedding extraction, and then clustering the speech regions in embedding space~\cite{wang2018speaker, 8937487park, 9053952moni}. Recently there has been an increase in end-to-end neural speaker diarization systems~\cite{huang2020speaker}. 
Specific to diarization in TV shows, previous methods have used visual patterns among the shots~\cite{bost2015audiovisual}, face clustering, and talking face detection~\cite{bredin2016improving, chung2020spot} to complement audio-driven speaker embeddings. 
Large-scale audio-visual diarization datasets around TV shows~\cite{chung2020spot} and feature films~\cite{xu2021ava} have also emerged, involving a semi-automatic annotation process.

%% file: methods.tex
\section{Methodology}
\subsection{Problem formulation}
Given a video, we acquire the set of speaker homogeneous speech segments $\{s_n\}, \forall n \in [1, N]$ and for each $s_n$, the corresponding temporally overlapping set of face-tracks, \mbox{$\{f_k^n\}, \forall k \in [1, K]$}. We define \emph{active speaker detection (ASD)} as the task of finding a source face-track (if any) for all speech segments $s_n$ from their respective sets of overlapping face-tracks $\{f_k^n\}$. We formulate \emph{speaker diarization (SD)} as a clustering task to cluster all the speech segments ${s_n}$ into $C$ clusters, $\{s_l^c\}, \forall l \in [1,L_c] \& c\in [1,C]$, where $L_c$ being the number of speech segments in the cluster $c$ and $C$ is the number of characters in the video, which in general is not known. We use oracle speaker homogeneous speech segments for all our experiments. 
\subsection{Active speaker detection (ASD)}
We use a two-stage method for ASD. We first derive active speaker information from the visual cues using a weakly supervised cross-modal framework~\cite{sharma2020cross}. We next construct audio-visual character profiles and impose a constraint that for a given active speaker instance $\{s_n, f_k\}$, source face-track $f_k$ for the speech segment $s_n$ belongs to the same character~\cite{sharma2022audio}. 
\subsubsection{Visual score}
\begin{figure}[tb]
    \centering
    \includegraphics[width=0.45\textwidth]{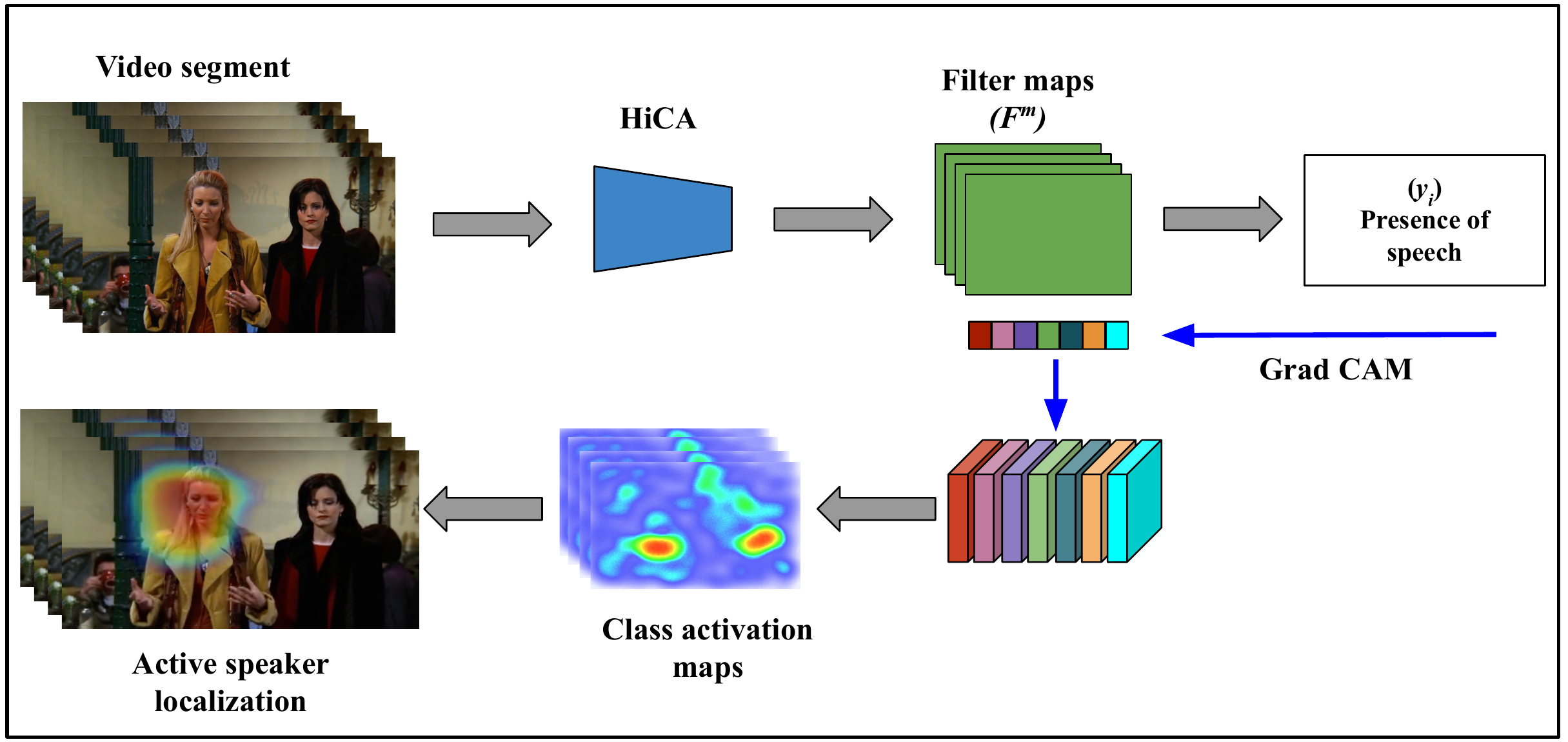}
    \caption{Active speaker localization using HiCA architecture and GradCAMs        ~\cite{sharma2020cross}.}
    \label{fig:asd}
\end{figure}

In our previous work, we proposed the Hierarchical Context Aware (HiCA) architecture~\cite{sharma2020cross}, trained for the task of the presence of speech in a video segment, and showed that the system could localize active speakers in video frames. We used class activation maps (CAMs) to localize the salient regions. Formally, for a video segment $v_i$, we compute a spatio-temporal ($3D$) feature map $F^m$ at the last layer of the HiCA network, with $m$ filters, and the softmax prediction score for the presence of speech, $y_i$. We then compute the CAM, $M$ following equaiton~\ref{eq:cams} with $Z$ being the averaging factor. Overview of the system is shown in Figure~\ref{fig:asd}.
\begin{align}
\label{eq:cams}
    \alpha_m &= \frac{1}{Z}\Sigma_i\Sigma_j\Sigma_k\frac{\partial {\hat{y_i}}}{\partial{F^{m}_{ijk}}} &
    M &= ReLU(\Sigma_m\alpha_mF^m)
\end{align}
For a face-track, $f_k$, temporally overlapping with speech segment, $s_n$, we compute a visual active speaker score, ($\text{VAS}_k$), by ROI pooling the relevant CAMs. We ROI pool the CAMs for each frame, $t$,  in the face-track that overlaps with $s_n$ in time and compute an aggregate mean over the frames, with $M_t$ denoting the CAMs for frame $t$ and $\{(x_1^t,y_1^t), (x_2^t,y_2^t)\}$ denotes the face bounding box in frame $t$. 
\begin{align}
\label{eq:vas}
    \text{VAS}_k &= \frac{1}{Z}\Sigma_f\Sigma_{x,y}M_t[y_1^t:y_2^t, x_1^t:x_2^t]
\end{align}

\subsubsection{Iterative profile matching}
In~\cite{sharma2022audio} we introduced a strategy constructing audio-visual character profiles to complement the visual-only active speaker detection, which takes advantage of the fact that for a potential active speaker instance $\{s_n, f_k\}$ if $s_n$ is a speech of one character $f_k $ should be the face of the same character. We initiate with a set of high confidence instances (HCI) of face-speech associations and cluster them to construct audio-visual character profiles $\{F_c, S_c\}$. $F_c \equiv \{f_k^c\}$ denotes the face instances clustered into character $c$ (the real identity and number of characters are unknown), and $S_c \equiv \{s_k^c\}$ denotes the corresponding speech instances for cluster $c$. 

Using the constructed audio-visual character profiles, we compute a profile matching score (PMS)~\cite{sharma2022audio} for all other instances as $P(s_n \rightarrow f_k )$, signifying the confidence of speech segment $s_n$ and face-track $f_k$ belonging to the same character profile. We update the set of HCI with the high scored active speaker instances (computed using a combination of VAS and PMS) and repeat the process until no more instances, scored high enough, to be added to HCI. At inference time we score any potential active speaker instance $\{s_n, f_k\}$ as $\text{VAS}_k +  \text{PMS}_{nk}$.

\subsection{Speaker diarization}
\label{subsec:SD}
The diarization objective is to cluster all instances of speech segments $s$ in the video such that speech segments coming from one character are grouped. Subsequent to active speaker detection, we have pairs of speech and associated face, $\{s_n, f_n\}$ along with speech instances $s_b$ which have speakers in the background; thus, no face associated. For audio-based diarization baseline, we use all the speech instances $s_n \cup s_b$ and cluster them using spectral clustering, post applying a diarization specific refinement sequence and estimating the number of speakers as suggested in~\cite{wang2018speaker}.

For vision-based clustering, we first filter the obtained set of speech-face associations $\{s_n, f_n\}$ based on the confidence of the association to remove the noisy instances. 
\begin{align}
A \equiv \{s_n, f_n\}: \text{VAS}_n + \text{PMS}_n > \tau    
\end{align}
We employ a density based clustering algorithm DBSCAN~\cite{DBSCAN}, which does not require knowledge of the number of clusters,  to cluster all the face-track instances $f_n \in A$ to obtain $F_c \equiv \{f_k^c\}, c\in [1,C]$, where $C$ is the number of clusters. DBSCAN also provides a list of instances which were not clustered and marked as noisy samples. Next we form clusters among speech segments $s_n \in A$, using the available associations $\{s_n, f_n\}$:
\begin{align}
    S_c \equiv \{s_n^c\}: \exists \{s_n^c, f_n^c\} \in A 
\end{align}
In order to assign cluster labels to the remaining speech segments (no face segments, noisy speech-face association, marked noisy by DBSCAN), we compute their cosine distance in speaker embeddings space from all the clustered points. We assign a cluster label to each instance $s_q$, the one that shows the least average distance, as shown in equation~\ref{eq:cluster_assign}, where $\{\cdot\}$ represents dot product and $\|.\|$ represents $l_2$ norm.
\begin{align}
\label{eq:cluster_assign}
\arg \min_{c} \frac{1}{|S_c|}\sum_n \frac{s_q \cdot s_n^c}{\|s_q\|\|s_n^c\|}    
\end{align}

%% file: Experiments.tex
\section{Experiments}
We use a subset of VPCD~\cite{2021VPCD}, consisting of 7 episodes from season 3 of the Friends TV show, for  evaluation purposes. VPCD provides all the face tracks appearing in the video frames and the character identities, and it also provides the speech segments for each character. It additionally consists of extracted face-embeddings for all the annotated faces and speaker embeddings for all the speech segments. We use the speaker homogeneous voices segments directly from the VPCD and use the provided face and speaker embeddings for our setup. 
\subsection{Active speaker detection performance}
The employed active speaker detection system predicts a corresponding face to each ground truth speech segment. We compute the accuracy of this system as the ratio of the duration of the correctly predicted faces to the total duration of the available speech segments. For a CAMs-only system, for a speech segment we assign the face-track with a maximum VAS score, among all the temporally overlapping face-tracks, as an active speaker. When combined with the iterative profile matching strategy, we use VAS + PMS score as the criterion. In Table~\ref{tab:asd} we compare the performance under those two scenarios and show that the profile matching strategy consistently improved the ASD performance for all the videos. 
\begin{table}[tb]
\centering
\caption{Active speaker detection performance improvement with iterative profile matching strategy.}
\label{tab:asd}
\resizebox{0.38\textwidth}{!}{
\begin{tabular}{ccc}
\hline
\begin{tabular}[c]{@{}c@{}}Video Name / \\ Approach\end{tabular} &
  \begin{tabular}[c]{@{}c@{}}CAMs only\\ (accuracy)\end{tabular} &
  \begin{tabular}[c]{@{}c@{}}CAMS + Iterative Profile\\ Matching (accuracy)\end{tabular} \\ \hline
Friends s03e17 & 0.69 & 0.76 \\ \hline
Friends s03e01 & 0.69 & 0.79 \\ \hline
Friends s03e02 & 0.65 & 0.75 \\ \hline
Friends s03e03 & 0.76 & 0.81 \\ \hline
Friends s03e04 & 0.77 & 0.81 \\ \hline
Friends s03e05 & 0.76 & 0.83 \\ \hline
Friends s03e06 & 0.77 & 0.84 \\ \hline
\end{tabular}%
}
\end{table}
\subsection{Speaker diarization performance}
\begin{figure*}
    \centering
    \includegraphics[width=\textwidth,keepaspectratio]{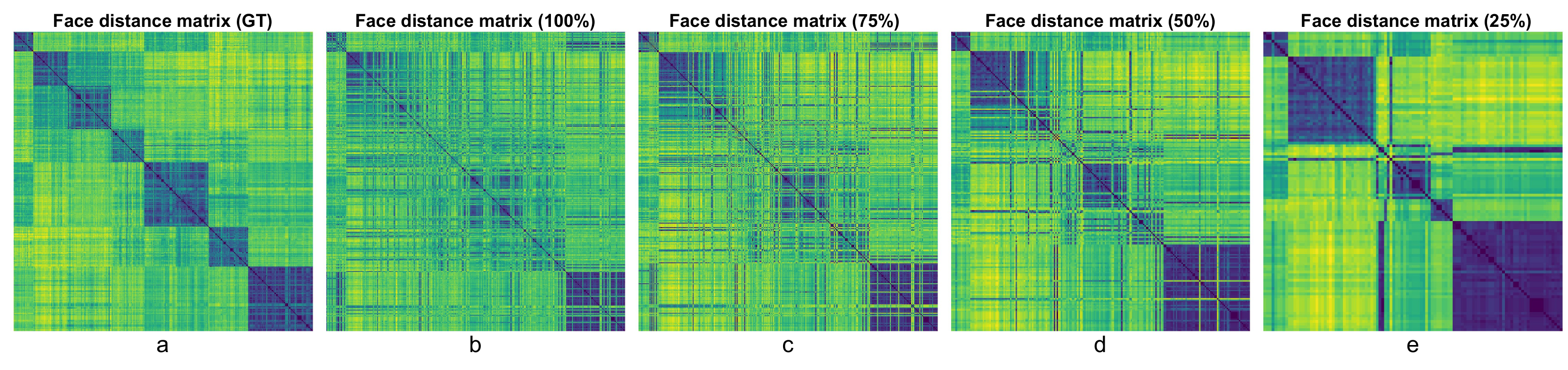}
    \caption{Distance matrices for the speech-face associations, using cosine distance among face-track embeddings, for different sets of speech-face instances. Selecting all (b), top 75\%(c), 50\%(d) and 25\%(e) samples in order of ASD scores. }
    \label{fig:system_vad}
\end{figure*}

As mentioned in section~\ref{subsec:SD}, before performing the face clustering for speaker diarization, we filter the obtained speech-face associations based on the confidence of the active speaker detection system. We present speaker diarization performance for four such filtered sets, containing top 25\%, 50\%, and 75\% of all the instances, ranked based on ASD score ($\text{VAS}+ \text{PMS}$) and one set containing all the samples (100\%). We compare the performance against the audio-based speaker diarization~\cite{wang2018speaker}. We report the diarization error rate (DER) as the metric to compare the SD performance, tabulated in Table~\ref{tab:DER_system}.

\begin{table}[tb]
\centering
\caption{Speaker diarization performance, DER (lower is better), using face-tracks compared against audio-only system.}
\label{tab:DER_system}
\resizebox{0.45\textwidth}{!}{%
\begin{tabular}{cccccc}
\hline
\begin{tabular}[c]{@{}c@{}}Video Name / \\ Top{x\%} ASD\end{tabular} &
  \begin{tabular}[c]{@{}c@{}}100\%\\ face-tracks\end{tabular} &
  \begin{tabular}[c]{@{}c@{}}75\%\\ face-tracks\end{tabular} &
  \begin{tabular}[c]{@{}c@{}}50\%\\ face-tracks\end{tabular} &
  \begin{tabular}[c]{@{}c@{}}25\%\\ face-tracks\end{tabular} &
  \begin{tabular}[c]{@{}c@{}}100\% \\ speech segments\end{tabular} \\ \hline
Friends s03e17 & 0.199          & 0.17          & \textbf{0.13} & 0.24          & 0.17   \\ \hline
Friends s03e01 & 0.27           & \textbf{0.16} & 0.2            & 0.27          & 0.27 \\ \hline
Friends s03e02 & 0.219          & 0.19        & 0.21          & \textbf{0.19} & 0.23   \\ \hline
Friends s03e03 & \textbf{0.18} & \textbf{0.18}         & 0.19           & 0.23          & 0.26  \\ \hline
Friends s03e04 & 0.31           & 0.3           & \textbf{0.13}  & 0.4           & 0.21   \\ \hline
Friends s03e05 & \textbf{0.16}  & 0.2           & 0.17           & 0.36         & 0.3    \\ \hline
Friends s03e06 & 0.21           & 0.24          & \textbf{0.17}  & 0.25          & 0.26   \\ \hline
\end{tabular}
}
\end{table}

We observe that the face-clustering-based diarization outperforms the audio-based framework for all the videos, except when we use just 25\% of samples for clustering. The relatively lower performance, in this case, can be attributed to the fact that 25\% of samples may not have enough information for all the characters. Most of the videos perform relatively worse while using 100\% of the samples than in the case with a lower number of high confidence points. This suggests that filtering of ASD results is a crucial step to remove noisy classifications before clustering them. 

To further understand the effects and dynamics of filtering the ASD results, in Figure~\ref{fig:system_vad} we show the distance matrices for the selected speech-face associations (for episode17, chosen arbitrarily), post-filtering, in the set $A$ (as described in section~\ref{subsec:SD}) for different setups. We compute the distances in the face domain, using the cosine distances between the face-track embeddings. In Figure~\ref{fig:system_vad}a, we show the ideal case scenario, with perfect ASD output (obtained using the ground truth). For visualization purposes, we grouped the instances for each character (using ground truth identities for speech segments) and used the same grouping for rest of the cases for easier comparison. In Figure~\ref{fig:system_vad}b, we show the distance matrix for all the speech-face associations obtained from the ASD system. The incorrect predictions are evident in the form of visible noise in the distance matrix compared against the ideal case scenario in Figure~\ref{fig:system_vad}a. In Figure~\ref{fig:system_vad}c,d,e we show the distance matrices of the filtered sets and observe that amount of noise decreases and the cluster patterns become more apparent. This results in improved SD performance with the filtered set, as shown in Table~\ref{tab:DER_system}, for the majority of videos. Although in Figure~\ref{fig:system_vad}e, showing the distance matrix for a smaller set of 25\% of all the samples, we notice prominent clusters, the number of data points drastically decreased, impacting the distribution of characters among them. Inadequate distribution of characters among the data points explains the drop in performance for the set containing 25\% samples, as reported in Table~\ref{tab:DER_system}.
\begin{figure}[tb]
    \centering
    \includegraphics[width=0.4\textwidth,keepaspectratio]{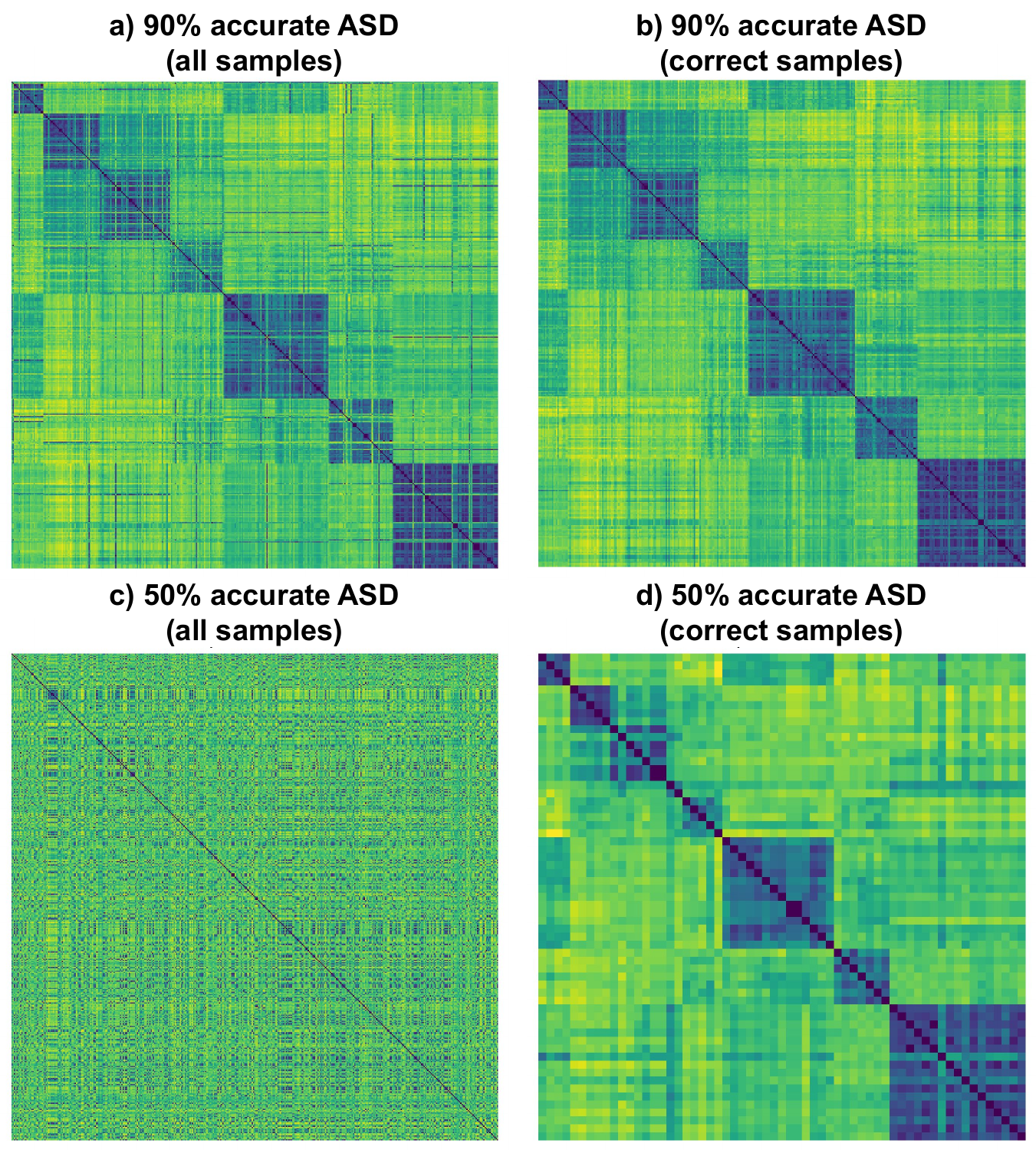}
    \caption{Distance matrices for simulated ASD output. a) and c): Simulated output with all samples. b) and d):s output with just the correct samples.}
    \label{fig:simulated}
\end{figure}

We note from the above discussion that the performance of ASD has a direct impact on the performance of SD. We simulated $k\%$ accurate ASD outputs, where $k \in [100, 90, 80, 70, 60, 50]$, using the available ground truth to study the same further. We randomly select $(100-k)\%$ of all the ground truth speech-face associations and shuffle their face instances among them, thus creating incorrect speech-face pairs. We perform speaker diarization using the simulated ASD outputs in two manners: i) using all the samples, including the incorrect ASD predictions, and ii) using just the correct ASD samples (we keep track of the correct samples while simulating).

\begin{table}[tb]
\centering
\caption{Variation in speaker diarization performance (using face clustering) with the quality of active speaker detection.}
\label{tab:simulated}
\resizebox{0.45\textwidth}{!}{%
\begin{tabular}{ccccccc}
\hline
\begin{tabular}[c]{@{}c@{}}Video name /\\ simulated accuracy\end{tabular} & 100\% & 90\% & 80\% & 70\% & 60\% & 50\% \\ \hline
s03e17 (all samples)     & 0.06 & 0.13 & 0.26 & 0.43 & 0.60 & 0.71 \\ 
s03e17 (correct samples) & 0.06 & 0.06 & 0.07 & 0.08  & 0.11 & 0.16 \\ \hline
s03e01 (all samples)     & 0.14 & 0.21 & 0.33 & 0.46  & 0.61 & 0.72 \\ 
s03e01 (correct samples) & 0.14 & 0.13 & 0.15 & 0.12 & 0.15 & 0.26 \\ \hline
s03e02 (all samples)     & 0.16 & 0.23 & 0.33 & 0.46  & 0.61 & 0.72 \\ 
s03e02 (correct samples) & 0.16 & 0.15 & 0.16 & 0.15  & 0.12 & 0.17 \\ \hline
\end{tabular}%
}
\end{table}

In Table~\ref{tab:simulated}, we report the DER for 3 randomly selected videos, using the two strategies, averaged over 5 runs. It is highly prominent that for when using all the samples, the speaker diarization performance drastically decreases as the accuracy of ASD decreases. It can be attributed to the introduction of more noisy speech-face associations as the accuracy for ASD falls. In Figure~\ref{fig:simulated} we show the distance matrices (simulated speech-face associations in terms of face-track embeddings) for two extreme values of $k$: 90\% and 50\%. The introduction of noise is visible in the 50\% accuracy case (figure~\ref{fig:simulated}c). 

In the case of speaker diarization using just the correct ASD instances, the DER remains nearly stagnant for higher accuracy but falls for lower accuracy (50\% case.) It can be attributed to the inadequate distribution of the samples across the characters in the video. The change in distribution can be noted in Figure~\ref{fig:simulated}b and d as the number of samples decreases for the lower accuracy case. Comparing the performance (while using just the correct samples) against the audio-based diarization performance (mentioned in Table~\ref{tab:DER_system}), we observe that even the lower accuracy cases outperform the audio-only systems.   